\documentclass[aps,prl]{revtex4}

\begin{document}
\title{{ Anomalous diffusion and anisotropic nonlinear Fokker-Planck equation}}
\author{E. K. Lenzi$^{1}$, R. S. Mendes$^{1}$ L. C. Malacarne$^{1}$ and L. R. da Silva$^{2}$}
\address{{$^{1}$ Departamento de F\'\i sica, Universidade Estadual de
Maring\'a, Av. Colombo 5790, 87020-900 Maring\'a-PR, Brazil \\
$^{2}$ International Center for Complex Systems and Departamento
de F\'{\i}sica, Universidade Federal do Rio Grande do Norte,
59078-970 Natal-RN, Brazil} }
\begin{abstract}
We analyse a bidimensional nonlinear Fokker-Planck equation by
considering an anisotropic case, whose diffusion coefficients
are $D_x \propto |x|^{-\theta}$ and $D_y \propto |y|^{-\gamma}$
with $\theta, \gamma \in {\cal{R}}$. In this context, we also
investigate two situations with the drift force
$\vec{F}(\vec{r},t)=(-k_{x}x, -k_y y)$. The first one is
characterized by $k_x/k_y=(2+\gamma)/(2+\theta)$ and the second is
given by $k_{x}=k$ and $k_{y}=0$. In these cases, we can verify an
anomalous behavior induced in different directions by the drift
force applied. The found results  are exact and exhibit, in terms of
the $q$-exponentials, functions which emerge from the Tsallis
formalism. The generalization for the $D$-dimensional case is
discussed.
\end{abstract}
\maketitle

\section{{ I - Introduction}}

The existence of the anomalous diffusion and its ubiquity has
motived the study of nonlinear Fokker-Planck equation (e.g.,
$\partial_{t} \rho = {\cal{D}}\nabla^{2}\rho^{\nu}$) due to its
broadness of applications. In fact, it has been employed in
several physical situations such as percolation of gases through
porous media ($\nu\geq 2$) \cite{buc6}, thin saturated regions  in
porous media ($\nu=2$)\cite{buc7}, a standard solid-on-solid model
for surface growth ($\nu=3$), thin liquid films spreading under
gravity ($\nu=4$) \cite{buc8}, among others \cite{buc10}. These
situations are essentially characterized by correlated anomalous
diffusion which has the second moment finite, in contrast with the
Levy distribution \cite{levy}. In recent works the nonlinear
Fokker-Planck equation, present here, has been analysed in one
dimension by considering linear force \cite{1a.1} and absorption
terms \cite{1e}. In \cite{renio} a spatial dependence was
incorporated in the diffusion coefficient, and in \cite{ct2000}
the nonlinear fractional Fokker-Planck was investigated in the
absence of drift.  The escape time, or mean first passage time,
has also been studied by employing a nonlinear Fokker-Planck
equation, leading eventually to a generalization of the Arrhenius
law \cite{celia}. Other aspects are analysed in \cite{outros}.
All these cases analyse the nonlinear Fokker-Planck equation in
one dimension or in $D$-dimension in an isotropic medium. However,
a careful analysis of the anisotropic case (${\cal D}_i$ $\neq$
${\cal D}_j$ and $F_i$$\neq$$F_j$, where ${\cal {D}}_i$ and $F_i$
are the diffusion coefficient and the external force) has not been
performed. Thus, it is desirable to study the above results
for the nonlinear Fokker-Planck equation in order to cover the
anomalous diffusive processes characterized by an anisotropic
medium. In this direction, we basically focus our attention on the
following anisotropic nonlinear Fokker-Planck equation:
\begin{widetext}
\begin{eqnarray}
\frac{ \partial \rho}{\partial t} = {\cal D}_x \frac{\partial }
{\partial x} \left \{ |x|^{-\theta} \frac{\partial
\rho^{\nu}}{\partial x} \right \} +  {\cal D}_y \frac{\partial
}{\partial y} \left \{ |y|^{-\gamma} \frac{\partial
\rho^{\nu}}{\partial y} \right \} - \nabla \cdot \left(
\vec{F}(\vec{r},t)\rho \right) \;\; , \label{anisotropico}
\end{eqnarray}
\end{widetext}
where $\vec{F}(\vec{r},t)$ is the external force
$\vec{F}(\vec{r},t)$$=$$(-k_{x} x, -k_{y} y )$ with $\theta$ and $
\gamma \in \cal{R}$. It can be verified that $\int_{-\infty}
^{\infty} d^{2}{r} \; \rho(\vec{r},t)$ is time independent (hence,
if $\rho$ is normalized at $t= 0$, it will remain so for ever).
Indeed, if we write the equation in the form $\partial _t \rho =
\nabla \vec{{\cal {J}}}$ and assume the boundary conditions
$\vec{{\cal {J}}}(\pm \infty,t) \rightarrow 0$, it can be shown
that $\int_{-\infty}^{\infty} d^{2}{r}\;\rho(\vec{r},t)$ is a
constant of motion.  Note that the cases mentioned above
\cite{1a.1,1e,renio,ct2000,outros}can essentially  be obtained
from Eq.(\ref{anisotropico}) by an adequate choice of the
parameters $\theta$, $\gamma$ and $\nu$. In particular,
Eq.(\ref{anisotropico}) reminds us of the equation used to describe
diffusion on fractals \cite{procaccia} and recovers, by a suitable
choice of ${\cal{D}}_x$, ${\cal{D}}_y$ and $\nu$, the special case
$\partial_t\rho=\overline{{\cal D}}_x
\partial_x\partial_x\ln(\rho)$, which emerges in plasma physics and
in the central limit approximation to Carleman\"us model of the
Bolztmann equation\cite{philip1}.

In this work, we first investigate the solutions of
Eq.(\ref{anisotropico}) by considering the absence of drift and
after we analyze the effect produced by the external force
$\vec{F}(\vec{r},t)=(-k_{x}x,-k_{y}y)$. In the last situation, the
external force is analyzed in two cases, the first one is
characterized by $k_x/k_y=(2+\gamma)/(2+\theta)$ and the second
one is given by $k_{x}=k$ and $k_{y}=0$. In addition, we show that
the presence of the nonlinearity produces a coupling among the
directions, in contrast with the usual Fokker-Planck equation.

\section{{ II- Anisotropic nonlinear Fokker-Planck equation}}

Before discussing the full case, we focus our attention on the
free case, i.e., in the absence of drift. Following the approach
employed in \cite{1a.1,renio,livro}, we can verify that the {\it
ansatz}
\begin{eqnarray}
\rho(\vec{r},t)=\left. \exp_q\left[ -\beta_x |x|^{2+\theta} -
\beta_y |y|^{2+\gamma} \right]\right / Z \label{solucao}
\end{eqnarray}
is the solution of Eq.(\ref{anisotropico}), with $\beta_x$, $\beta_y$
and $Z$ being time dependent parameters  to be determined and
$\nu=2-q$. The $\exp_{q}\left[ x\right]\equiv
[1+(1-q)x]^{\frac{1}{1-q}}$ is the $q$-exponential which arises from
the Tsallis formalism\cite{1a}. It is interesting to mention that
some properties of the renormalization group established in
\cite{RC}, such as $\rho(x,t)=\int_{-\infty}^{\infty} dy
\rho(\vec{r},t)$, are verified by Eq.(\ref{solucao}) as well as
the scaling dependence discussed in \cite{godenfeld}. In
particular, it is simple to verify that Eq.(\ref{solucao}) can be
obtained from the Tsallis entropy
\begin{eqnarray}
S_q=\frac{1-{\mbox {Tr}}\rho^q}{q-1}
\end{eqnarray}
by applying the maximum principle of entropy adequated constraints
taking into account, $\langle \langle x^{2+\theta} \rangle
\rangle=\langle \langle y^{2+\gamma} \rangle \rangle= const.$ with
$\langle \langle \cdots \rangle \rangle =\int d^{2}r
[\rho(\vec{r},t)]^q (\cdots) /\int d^{2}r [\rho(\vec{r},t)]^q$
(see also \cite{livro}). In order to obtain an explicit form for
$\beta_x$, $\beta_y$ and $Z(t)$ we substitute the {\it ansatz}
given by Eq.(\ref{solucao}) into Eq.(\ref{anisotropico}).
This replacement leads to the equations:
\begin{eqnarray}
\frac{1}{Z}
\frac{\partial Z}{\partial t} &=&\nu (2+\theta) {\cal
D}_x
\beta_x Z^{1-\nu} + \nu (2+\gamma) {\cal D}_y \beta_y Z^{1-\nu}\nonumber \; ,\\
\frac{1}{\beta_x}\frac{\partial \beta_x }{\partial t} &=&-\nu
(2+\theta)^2 {\cal D}_x\beta_x Z^{1-\nu}\; , \nonumber \\
\frac{1}{\beta_y }\frac{\partial \beta_y }{\partial t} &=&-\nu
(2+\gamma)^2 {\cal D}_y \beta_y Z^{1-\nu} .\label{SISTEMA}
\end{eqnarray}
From the above system of equations we can deduce the relation
\begin{equation}
\frac{Z(t)}{Z(t_{0})}
\left(\frac{\beta_x(t)}{\beta_{x}(t_{0})}\right)^{\frac{1}{2+\theta}}
\left( \frac{\beta_y(t)}{\beta_{y}(t_{0})
}\right)^{\frac{1}{2+\gamma}} =1 \;\; , \label{relacao}
\end{equation}
where $\beta_{x}(t_{0})$, $\beta_{y}(t_{0})$ and $Z(t_{0})$ are
constants. Now, employing Eq.(\ref{relacao}) in Eq.(\ref{SISTEMA})
and using the relation
\begin{eqnarray}
\beta_x(t)=\left( \frac{2+\gamma}{2+\theta} \right)^2 \frac{{\cal
D}_y}{{\cal D}_x}\beta_y(t) \label{betax}
\end{eqnarray}
we verify that
\begin{widetext}
\begin{eqnarray}
\beta_y(t) =\left \{ \left( 1-\varepsilon\right)(2+\gamma)^{2}\nu
{\cal D}_y {\cal C}^{1-\nu} t  \right
\}^{-\frac{1}{1-\varepsilon}} \;, \label{beta}
\end{eqnarray}
\end{widetext}
where ${\cal C}=Z(t_{0}) \beta_{x}^{\frac{1}{2+\theta}} (t_{0})
\beta_{y}^{\frac{1}{2+\gamma}}(t_{0}) [(2+\gamma)^2{\cal
D}_y/((2+\theta)^2{\cal D}_x)]^{-\frac{1}{2+\theta}}$ and
$\varepsilon=(4+\theta+\gamma) (1-\nu)/((2+\theta)(2+\gamma))$. To
find $Z(t)$, we may employ Eqs.(\ref{relacao}), (\ref{betax}) and
(\ref{beta}).

We may obtain further information from the the averages and
dispersion relations in order to investigate which combination of
$\theta$, $\gamma$ and $\nu$ leads us to a sub, normal or super
diffusion. Thus, for simplicity, we analyse the behavior of
$\langle x^{2} \rangle$, $\langle y^2 \rangle$ and $\langle
|r|^{2} \rangle$. Let us now obtain $\langle x^2 \rangle$. By
using the definition of average we can show that
\begin{widetext}
\begin{eqnarray}
\left\langle x ^{2}\right\rangle  &=&\frac{\int d^{2}r x^{2}
\rho({\vec{r}},t)}{\int d^{2}r\rho (
{\vec{r}},t)} \nonumber \\
&=&\frac{1}{\beta_x^{\frac{2}{2+\theta}}(t)} \left\{
\begin{array}{cc} \frac{\Gamma \left( \frac{3}{2+\theta}\right)
\Gamma \left( \frac{1}{q-1}-\frac{3}{2+\theta}-\frac{1}{2+\gamma}
\right)}{(q-1)^{\frac{2}{2+\theta}}\Gamma\left(\frac{1}{2+\theta}
\right)\Gamma \left(
\frac{1}{q-1}-\frac{1}{2+\theta}-\frac{1}{2+\gamma} \right)}
& q>1  \\
\frac{\Gamma \left( \frac{3}{2+\theta} \right) \Gamma
\left(\frac{2-q}{1-q}+\frac{3}{2+\theta}+\frac{1}{2+\gamma}\right)
} {(1-q)^{\frac{2}{2+\theta}}\Gamma\left( \frac{1}{2+\theta}
\right)\Gamma \left(\frac{2-q}{1-q}+ \frac{3}{2+\theta}+
\frac{1}{2+\gamma} \right)} \text{ } & q<1
\end{array}
\right. \; .
\label{MEDIA}
\end{eqnarray}
\end{widetext}
The behavior of Eq.(\ref{MEDIA}), i.e., sub, normal or
super-diffusive, depends on the relation
$\theta-((4+\theta+\gamma)(1-\nu))/(2+\gamma)$ to be less, equal
or greater than zero. In the same way, it can be shown that
$\langle y^2 \rangle$ $\propto 1/\beta_y^{\frac{2}{2+\gamma}}(t)$
and $\langle |\vec{r}|^2 \rangle \propto
1/\beta_x^{\frac{2}{2+\theta}}(t)+\overline{{\cal {C}}}/
\beta_y^{\frac{2}{2+\gamma}}(t)$, where $\overline{{\cal {C}}}$ is
a constant. In particular, it is interesting to note that $\langle
|\vec{r}|^2 \rangle$ may exhibit two regimes where the last one
presents a diffusion that is faster than the first one, see
Fig.(\ref{fig}). Similar behavior for the second moment is found
in \cite{anisotropico1} by considering the anomalous walker
diffusion through composite systems.

Now, let us analyse the implications obtained by considering
the drift $\vec{F} (\vec{r},t)=(-k_x x,-k_y y)$ with
$k_x/k_y=(2+\gamma)/ (2+\theta)$. The drift incorporated in
Eq.(\ref{anisotropico}) only leads us to a modifications in
(\ref{SISTEMA}) altering the time dependence on $\beta_y$ as
follows:
\begin{eqnarray}
\beta_y=\left \{ \frac{(2+\gamma)\nu {\cal D}_y}{k_y} {\cal
C}^{1-\nu} \left( 1-e^{(1-\varepsilon)(2+\gamma)k_yt} \right)
\right \}^{-\frac{1} {1-\varepsilon}}  \; .
\end{eqnarray}
Note that the limit $k_y\rightarrow 0 $ recovers (\ref{beta}). Let
us now consider the case $k_{x}=k$ and $k_{y}=0$, with $\gamma$
and $\theta$ being arbitrary, the equations for the time dependent
parameters do not have explicit solutions. In fact, we may only
obtain an implicit solution as follows:
\begin{widetext}
\begin{eqnarray}
\beta_{x}(t)&=&\left [
\beta_{x}^{\frac{q-3-\theta}{2+\theta}}(t_{0}) e^{-(1+\theta+\nu)k
t} +(1+\theta+\nu)(2+\theta) \nu{\cal
{D}}_{x}{\cal{C}}_{1}^{1-\nu}e^{-(1+\theta+\nu)k t}
\int_{t_{0}}^{t} d\overline{t} \; e^{(1+\gamma+\nu)k \overline{t}}
\beta_{y}(\overline{t})^{\frac{-1+\nu}{2+\gamma}} \right
]^{-\frac{2+\theta} {1+\theta+\nu}}\;,  \nonumber \\
\beta_{y}(t)&=&\left [
\beta_{y}^{\frac{q-3-\gamma}{2+\gamma}}(t_{0})
 +(1+\gamma+\nu)(2+\gamma) \nu{\cal
{D}}_{y}{\cal{C}}_{1}^{1-\nu}\int_{t_{0}}^{t} d\overline{t} \;
\beta_{x}(\overline{t})^{\frac{-1+\nu}{2+\theta}} \right
]^{-\frac{2+\gamma} {1+\gamma+\nu}}\; , \label{nl-1}
\end{eqnarray}
\end{widetext}
with ${\cal{C}}_{1}=Z(t_{0})\beta_x^{\frac{1}{2+\theta}} (t_{0})
\beta_{y}^{\frac{1}{2+\gamma}}(t_{0})$.

From the developments made above we can verify an interesting fact
which does not appear in the usual case, the external force applied
only in one direction, for example, $x$-direction, manifests itself
in all the other directions. This fact may be seen in the time dependence
parameters which in this situation are coupling; see the systems of
equation (\ref{nl-1}).
In particular, the result that emerges from the above system of
equations (\ref{nl-1}) by considering the second order of the
perturbation theory near $t_{0}$ for $\beta_{y}(t)$, as follow:
\begin{widetext}
\begin{eqnarray}
\beta_{y}(t)&=&\left [
\beta_{y}^{\frac{q-3-\gamma}{2+\gamma}}(t_{0})
 +(1+\gamma+\nu)(2+\gamma) \nu{\cal
{D}}_{y}{\cal{C}}_{1}^{1-\nu}\int_{t_{0}}^{t} d\overline{t} \;
\left( \beta_{x}^{\frac{q-3-\theta}{2+\theta}}(t_{0})
e^{-(1+\theta+\nu)k(\overline{t}-t_{0})} \right. \right. \nonumber \\
&+& \left. \left. \frac{(2+\theta){\cal{D}}_{x}
\nu}{k_{x}}{\cal{C}}_{1}^{1-\nu}\beta_{y}^{\frac{1-q}{2+\gamma}}(t_{0})
\left(1-e^{-(3+\theta-q)k(t-t_{0})}\right)
\right)^{\frac{1-\nu}{3+\theta-q}} \right]^{-\frac{2+\gamma}
{1+\gamma+\nu}} \; ,  \label{nl-2}
\end{eqnarray}
\end{widetext}
shows that the external force $\vec{F}=(-k x,0)$ applied only in
the $x$ direction modifies the $y$ direction where we have no
drift. In this direction, a simple way to identify, without
approximation, the effect of a external force, for the spatial
dependence, in an orthogonal direction is to analyse the stationary
solution. In particular,  for Eq.(\ref{anisotropico}), by
considering the drift
$\vec{F}(\vec{r},t)=(-\frac{d}{dx}V_{x}(x)),-\frac{d}{dy}V_{y}(y))$
and for simplicity $\theta=0$ and $\gamma=0$, the stationary
solution is given by
\begin{eqnarray}
\rho(\vec{r},t)= \left. \exp_q\left[-\beta_x V_x(x)-\beta_y
V_y(y)\right] \right / Z_q \; ,
\end{eqnarray}
with $\beta_{y}$ and $\beta_{y}$ obeying the relation
${\cal{D}}_{x}\beta_{x}={\cal{D}}_{y}\beta_{y}$and
\begin{eqnarray}
Z_q=\int \exp_q[-\beta_{x}V_{x}(x) -\beta_{y}V_{y}]\; .
\end{eqnarray}

\section{{III - Summary and Conclusions}}

In summary, we have worked the nonlinear Fokker-Planck  equation
in several situations by incorporating an anisotropic dependence
in the difusion coefficients and also in the external force. We
first analysed the anisotropic case with diffusion coefficients
dependent on the position variables and we found its solution,
Eq.(\ref{solucao}). When we incorporated a drift term with
$k_{x}/k_{y}=(2+\gamma)/ (2+\theta)$ we verified that
Eq.(\ref{solucao}) is the solution with a different dependence on time
for the parameters $\beta_{x}$, $\beta_{y}$ and $Z$. For the case
$k_{x}=k$ and $k_{y}=0$, we show that Eq.(\ref{solucao}) is also
the solution; however, the explicit time dependence on
the parameter $Z$, $\beta_{x}$ and $\beta_{y}$ is very
complex leading us to determine them as implicitly. In particular, we
have shown that an external force that has only one component when
applied to the system governed by Eq.(\ref{anisotropico}) changes
the behavior of all directions and consequently the time evolution
of the distribution, in contrast with the usual case. The
generalization of these results for the multidimensional case is
direct when we consider the absence of the drift. In the presence of
dritf, we may work the expression in order to recover the above
discussion for the bidimensional case. It is important to remark
here that the solutions found here can no be written as
$\rho(\vec{r},t)=\rho(x,t)\rho(y,t)$ and an external force applied
to the system modifies the behavior of the distribution in all
direction, in contrast with the usual case. Finally, we expect
that the results obtained here can be useful in the discussion of
the anomalous diffusion and its connection with the nonextensive
Tsallis thermostatistics.

\section*{{ Acknwonledgements}}

We thank CNPq (Brazilian agency) for partial financial support.

\begin{figure}
\caption{{This figure illustrates the behavior of  $\langle
|\vec{r}|^2 \rangle$ for typical values $\nu$, $\theta$ and
$\gamma$.}} \label{fig}
\end{figure}

\end{document}